\documentclass[times, review, 10pt]{elsarticle}
\biboptions{sort&compress} %

\usepackage{amssymb}
\usepackage{amsmath}

\usepackage{amsmath,amsfonts}
\usepackage{algorithmic}
\usepackage{algorithm}
\usepackage{array}
\usepackage[caption=false,font=normalsize,labelfont=sf,textfont=sf]{subfig}
\usepackage{textcomp}
\usepackage{stfloats}
\usepackage{url}
\usepackage{verbatim}
\usepackage{graphicx}
\usepackage{booktabs}
\usepackage{multirow}
\journal{Pattern Recognition}
\usepackage{setspace}
\begin{document}

\begin{frontmatter}



\title{CMADiff: Cross-Modal Aligned Diffusion for Controllable Protein Generation}

\author[inst1,inst2]{Changjian Zhou}
\author[inst1,inst2]{Xinyue Zhang}
\author[inst3]{Yuexi Qiu}
\author[inst4]{Jiaqing Zhang}
\author[inst5]{Heng-Da Cheng}
\author[inst1]{Jia Song\corref{cor1}}
\ead{songjia@neau.edu.cn}
\author[inst1]{Wensheng Xiang\corref{cor1}}
\ead{xiangwensheng@neau.edu.cn}
\cortext[cor1]{Corresponding authors.}
\affiliation[inst1]{organization={Key Laboratory of Agricultural Microbiology of Heilongjiang Province, Northeast Agricultural University},
            city={Harbin},
            postcode={150030}, 
            state={Heilongjiang},
            country={China}}
            
\affiliation[inst2]{organization={School of Intelligent Science and Engineering, Northeast Agricultural University},
            city={Harbin},
            postcode={150030}, 
            state={Heilongjiang},
            country={China}}

\affiliation[inst3]{organization={School of Software Technology, Zhejiang University},
            city={Ningbo},
            postcode={315100}, 
            state={Zhejiang},
            country={China}}
            
\affiliation[inst4]{organization={Sydney Smart Technology College, Northeastern University},
            city={Qinhuangdao},
            postcode={066004}, 
            state={Liaoning},
            country={China}}
            
\affiliation[inst5]{organization={Department of Computer Science, Utah State University},
            city={Logan},
            postcode={84322}, 
            state={Utah},
            country={USA}}
            
\begin{abstract}
AI-assisted protein design has emerged as a critical tool for advancing biotechnology, as deep generative models have demonstrated their reliability in this domain. However, most existing models primarily utilize protein sequence or structural data for training, neglecting the physicochemical properties of proteins. Moreover, they are deficient to control the generation of proteins in intuitive conditions. To address these limitations, we propose CMADiff here, a novel framework that enables controllable protein generation by aligning the physicochemical properties of protein sequences with text-based descriptions through a latent diffusion process. Specifically, CMADiff employs a Conditional Variational Autoencoder (CVAE) to integrate physicochemical features as conditional input, forming a robust latent space that captures biological traits. In this latent space, we apply a conditional diffusion process, which is guided by BioAligner, a contrastive learning-based module that aligns text descriptions with protein features, enabling text-driven control over protein sequence generation. Validated by a series of evaluations including AlphaFold3, the experimental results indicate that CMADiff outperforms protein sequence generation benchmarks and holds strong potential for future applications. The implementation and code are available at~\url{https://github.com/HPC-NEAU/CMADiff}.
\end{abstract}


\end{frontmatter}

\section{Introduction}
Proteins are essential organic macromolecules that perform a wide range of biological functions and play an indispensable role in molecular systems within living organisms \cite{1}. Moreover,  proteins are not restricted to those which could be found in nature, as synthesized proteins in the laboratory also exhibit biological functions. Therefore, natural proteins represent only a small fraction of the vast protein sequence space \cite{2}. Exploring this sequence space through the design of proteins to produce biologically functional proteins remains a daunting challenge \cite{3}. Currently, protein design paradigms can be broadly categorized into structure-based design and sequence-based design.
Despite the notable success of structure-based models in recent years \cite{4,5,6}, conventional protein generation methods still rely on the expression of sequences through genetic mechanisms, since proteins are ultimately encoded by their sequences. Consequently, sequence-based methods align more closely with real-world applications, offering greater practicality as a direct approach for protein generation. For example, Alamdari, et al. proposed EvoDiff \cite{7}, which integrated the principles of diffusion models to design biological plausible protein sequences. As demonstrated in the preceding literature, analogous sequence-based models have also exhibited proficiency in generating high-quality sequences \cite{8}. However, these models have only contributed to the methodology of unconditional generation, as they generate sequences without the presence of specific directives or constraints. These methods are time-consuming and labor-intensive in practical scenarios, since researchers have to manually screen the generated sequences to identify proteins with desired functions \cite{9,10}. This limitation underscores the need for more targeted approaches that are capable of designing proteins with specific functional properties.
To address the challenge of uncontrollability, Lin et al. proposed TaxDiff \cite{11}, which utilized  classification information of species to guide diffusion in protein generation. However, in practical applications, controlling protein generation based on taxonomy has its limitation because species alone could not specify the actual functions of proteins. Generally, protein design needs to be conditioned on functional characteristics, as the physicochemical properties of proteins essentially reflect their functional and structural features \cite{12}. Early research  has established that local biochemical signatures, such as hydropathy blocks, provide sufficient discriminative power to classify protein families and infer their biological roles \cite{A}.Currently, a large amount of knowledge has been organized in text formats to describe higher level functions of proteins, such as UniProt database \cite{13}, \textit{etc}. Furthermore, methods that leverage drug text descriptions alongside intrinsic chemical and structural information have been proven effective in small molecule drug discovery \cite{14,15,16}. Inspired by these factors, it is possible to take advantage of textual descriptions as conditions for generating models in similar domains. Therefore, we aim to further investigate the possibility of fusing conditional diffusion and physicochemical properties of proteins for generating proteins with specific functions.
\begin{figure}[htbp]
\centering
\includegraphics[width=3.5in]{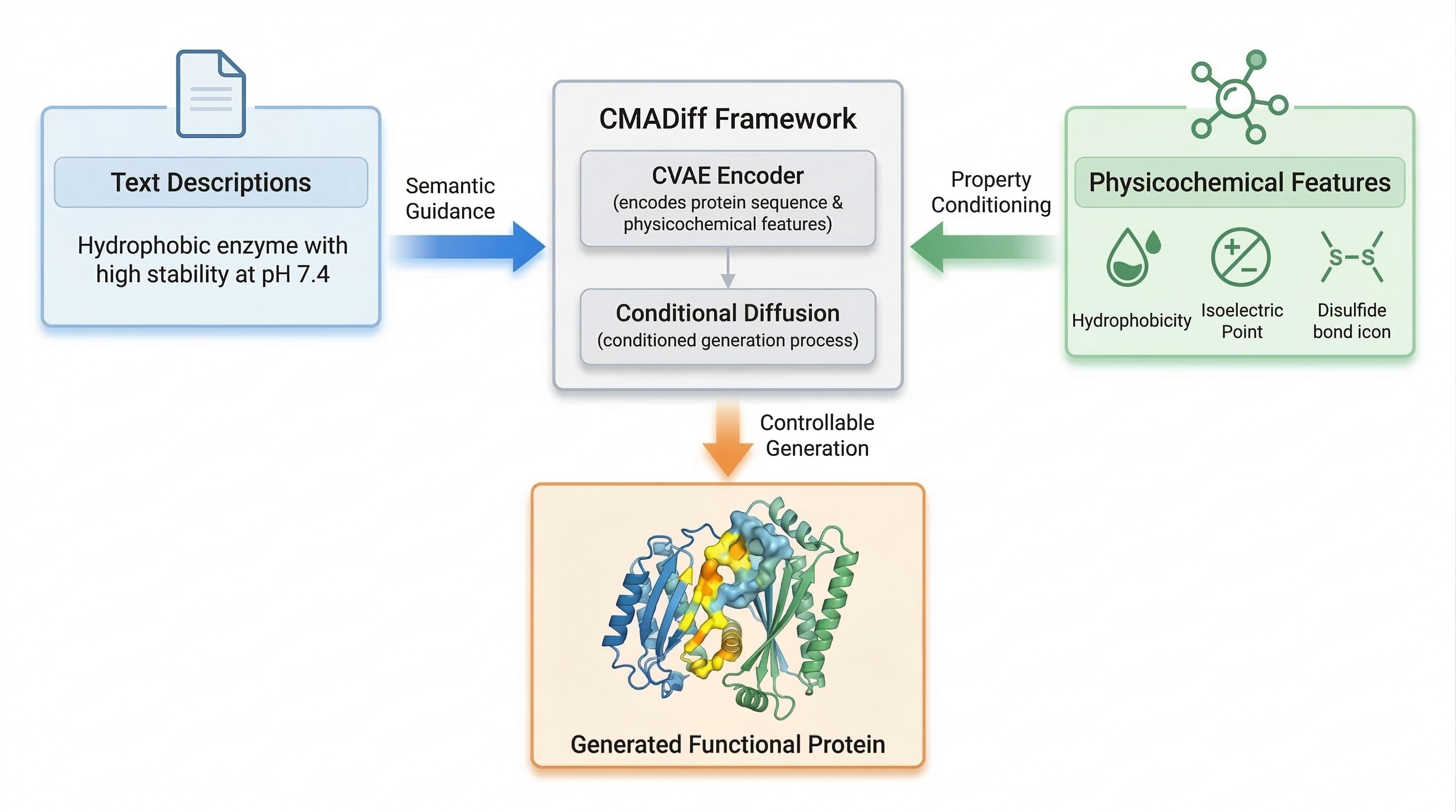}
\caption{Physicochemical properties and text descriptions guide functional protein design}\label{fig1}
\end{figure}
 Driven by the physicochemical properties of proteins and their corresponding functional text descriptions, we propose CMADiff, a property-driven framework for generating bioactive proteins with specified functions. As depicted in \textbf{Figure \ref{fig1}}, the objective of this study is to facilitate the generation of specific proteins through the use of CMADiff, where users provide textual descriptions and physicochemical data. CMADiff employs a conditional variational autoencoder (CVAE) \cite{cvae} to encode protein sequences and their physicochemical features into a latent space, ensuring robust representations of biological properties.Such conditional generative frameworks have demonstrated significant efficacy in similar domains. For instance, the Structure-aware CVAE  was proposed to optimize molecular properties by aligning structural topology within a variational framework \cite{SCVAE_paper}. The model incorporates both one-hot encoding and local physicochemical features of the sequences, while global physicochemical properties, averaged across the sequence, serve as conditional input to the CVAE. In order to enhance flexibility and provide fine-grained control over sequence generation, a conditional diffusion process is embedded into the latent space. For more intuitive text-driven protein generation via this process, we designed the BioAligner module. BioAligner aligns protein textual descriptions with their physicochemical properties. Therefore, enabling precise control over protein traits through textual conditions. The main contributions of this work are as follows:
 
$\bullet$~A novel protein sequence generation framework is proposed in this work. Not only can it generates high-scoring sequences with structural similarities to natural proteins, but also novel and biological plausible sequences that have not yet been found in nature.

$\bullet$~To the best of our knowledge, CMADiff is the first model to utilize physicochemical features and text descriptions in protein generation tasks.

$\bullet$~ A BioAligner module is designed to bridge the gap between textual annotations and physicochemical features.

$\bullet$~The AlphaFold 3 is employed for evaluation, and the results indicated that CMADiff outperforms existing state-of-the-art generation models, demonstrating its great potential.

\section{Preliminary}
\subsection{Protein sequence representation}
A protein sequence \( S \) of length \( L \) is represented in Equation\eqref{eq:protein_sequence}.
\begin{equation}
S = \{s_1, s_2, \dots, s_L\}, \quad s_i \in A\label{eq:protein_sequence}
\end{equation}
where \( A \) is the set of 20 standard amino acids. To capture the semantic context of proteins, each text annotation \( T \) is encoded into a dense vector representation \( t \in \mathbb{R}^d \) using a pre-trained language model. 

\subsection{Physicochemical features}

Physicochemical properties provide essential biochemical context, augmenting sequence information for biologically meaningful generation.

\textbf{Local features:} For each amino acid \( s_i \) in the sequence \( S \), we define a local physicochemical feature vector \( f(s_i) \in \mathbb{R}^k \), where \( k \) is the number of physicochemical properties (e.g., hydrophobicity, charge, polarity). The full sequence of local features \( f \in \mathbb{R}^{L \times k} \) is then concatenated with the one-hot encoded representation of the protein sequence to form a joint local representation of \( S \).

\textbf{Global features:} The global physicochemical features \( \bar{f} \in \mathbb{R}^k \) represent an averaged vector of the physicochemical properties across the entire protein sequence, as profiled in Equation\eqref{2}. 
\begin{equation}
\bar{f} = \frac{1}{L} \sum_{i=1}^{L} f(s_i)\label{2}
\end{equation}

\( f(s_i) \) is the local physicochemical feature vector for the \( i \)-th amino acid in the sequence. These global features summarize the overall physicochemical characteristics of the sequence.

\subsection{Latent space representation}

The latent space serves as a compact, biologically meaningful representation that encodes both sequence-level and physicochemical information. The encoder maps both the protein sequence \( S \) and its local physicochemical features \( f \) (along with the global features \( \bar{f} \)) into a latent representation \( z \in \mathbb{R}^m \), where \( m \) is the latent dimension.
\begin{equation}
z = f_{\text{enc}}(S, f, \bar{f}) \label{3}
\end{equation}

Equation\eqref{3} ensures that \( z \) captures both local sequence information and global physicochemical context.

\subsection{Conditional generation}

To enable text-guided generation, the model leverages latent representations and textual conditions to synthesize biologically relevant protein sequences. Given a text condition \( t \), the diffusion model generates a new protein sequence \( S^* \) via reverse diffusion from a noise vector \( z_* \), as profiled in Equation\eqref{4}.
\begin{equation}
 S^* = f_{\text{decode}}(f_{\text{diffusion}}(z_*, t))\label{4}   
\end{equation}

\( z_* \) is a noise-perturbed latent variable sampled from a Gaussian distribution. \( f_{\text{diffusion}} \) is the denoising process, conditioned on \( t \), that refines \( z_* \) to reflect the biological context specified by \( t \).  \( f_{\text{decode}} \) reconstructs \( S^* \) using the refined latent representation and global features \( \bar{f} \) to ensure biological plausibility.

\section{Method}\label{sec3}

\begin{figure}[h]
\centering
\includegraphics[width=3.5in]{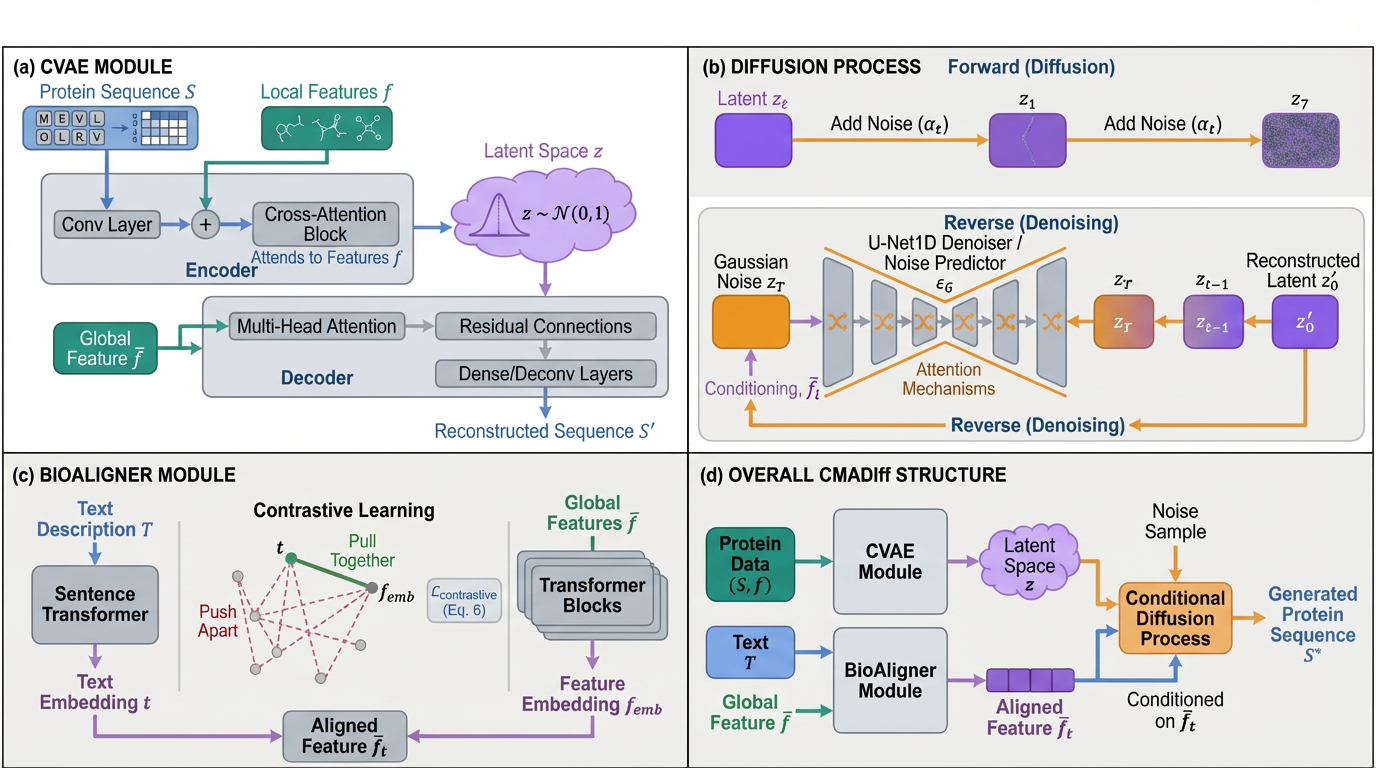}
\caption{ \textbf{CMADiff architecture.} \textbf{(a)} CVAE module: The  EncoderBlock combines convolution and cross-attention block \cite{attention} to focus on physicochemical features \cite{zhou2024transvae}, while the DecoderBlock uses multi-head attention with latent space and conditional information, enhanced by residual connections \cite{zhou2023rice}. \textbf{(b)} diffusion process:The diffusion process employs DDPM \cite{ddpm} with a U-Net1D-based noise predictor, incorporating attention mechanisms. \textbf{(c)} The BioAligner module uses contrastive learning between a Sentence Transformer for text and TransformerBlocks for physicochemical features. \textbf{(d)} The overall structure of CMADiff}\label{fig2}
\end{figure}

\subsection{Overall architecture}\label{subsec1}
As illustrated in \textbf{Figure \ref{fig2}}, CMADiff integrates physicochemical properties and semantic guidance into a diffusion-based protein sequence generation framework. \textbf{Figure \ref{fig2}(a)} demonstrates local and global physicochemical features (denoted as \( f \)) are encoded into the latent space (represented by \( z \)) by the CVAE module, while \textbf{Figure \ref{fig2}(b)} depict the U-Net1D-based diffusion process refines the noisy representation of \( z \). In \textbf{Figure \ref{fig2}(c)}, the BioAligner module aligns textual descriptions, denoted as \( T \), with physicochemical features, denoted as \( F \), to enable text-guided controllable generation.

\begin{figure}[h]
\centering
\includegraphics[width=3.5in]{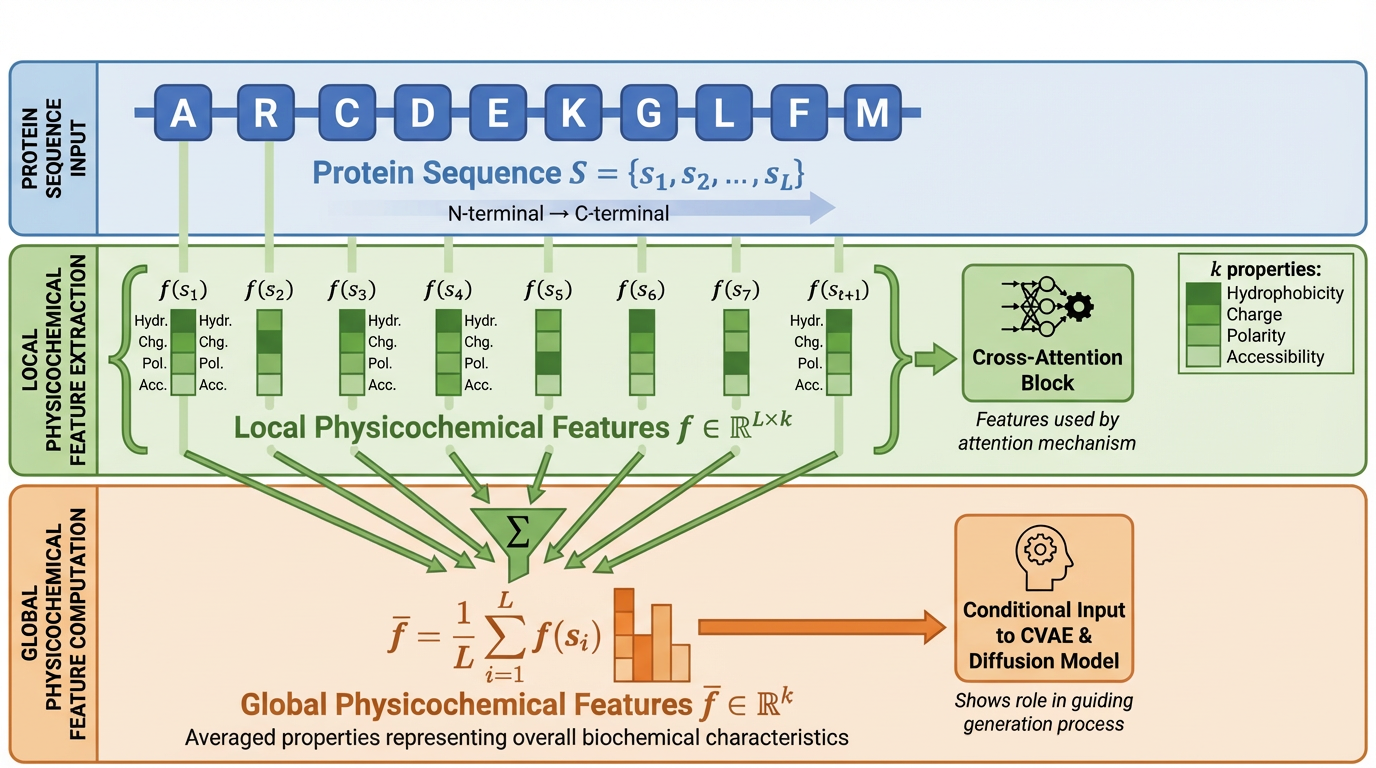}
\caption{Global and local physicochemical properties}\label{cvae}
\end{figure}

\subsection{Latent space in CVAE}\label{subsec2}
The CVAE framework integrates protein sequence information and physicochemical properties in a structured latent space, facilitating both local and global representation of proteins. Protein sequences are represented of one-hot encoded vectors \( S \in \mathbb{R}^{L \times 21} \), where \( L \) is the sequence length, and the 21 dimensions include the 20 standard amino acids plus padding. Local physicochemical features \( f \in \mathbb{R}^{L \times k} \) (such as hydrophobicity and polarity) are also included for each position in the sequence. To process these inputs, the model uses separate residual convolutional layers for both sequence and physicochemical features. These representations are then combined using a cross-attention mechanism and passed through a transformer encoder, which captures complex dependencies between the sequence and its physicochemical properties. This step generates the latent variables by computing the mean vector \( \mu \) and log-variance vector \( \log \sigma^2 \), which are used to sample the latent variables \( z = \mu + \sigma \cdot \epsilon \) via the reparameterization trick (\( \epsilon \sim \mathcal{N}(0, I) \)).

A global feature vector is generated by averaging the physicochemical features across the entire protein sequence. This vector encapsulates the overall biochemical properties of the protein, ensuring that the generated sequences adhere to global biological plausibility. This global representation serves as a conditional input to the CVAE decoder, influencing the generated protein sequence to meet higher-level biochemical criteria. As shown in \textbf{Figure} \ref{cvae}, the global feature is derived from the average of the physicochemical properties across the sequence, while local features correspond to position-specific properties that are integrated into the model using a cross-attention mechanism. 

The latent space is regularized to align with a standard Gaussian distribution \( \mathcal{N}(0, 1) \), promoting smooth interpolation and robust sampling. This is achieved by including a KL divergence term in the loss function \(\L_{\text{CVAE}}\), as stated in Equation\eqref{5}.
\begin{equation}
L_{\text{CVAE}} = \mathbb{E}_{q(z|S,f)} \left[ \|S - S'\|^2 \right] + \beta \cdot D_{\text{KL}}(q(z|S,f) \| \mathcal{N}(0,1)) \label{5}
\end{equation}

\( \beta \) controls the trade-off between reconstruction fidelity and KL divergence, ensuring the model learns both accurate sequence generation and latent space regularization.

The decoder takes the sampled latent variable \( z \) and the global physicochemical features \( \bar{f} \) as inputs. These are processed through dense layers, transformer decoders, and deconvolutions to generate the output protein sequence \( S' \), which aligns with both the input sequence and its corresponding physicochemical properties. By incorporating physicochemical properties, the model generates protein sequences that are biologically relevant and exhibit desired biochemical traits. The latent space supports smooth transitions. This allows flexible generation of sequences with different properties. For robust and accurate protein design, the use of both local and global representations ensures that the generated sequences closely match real-world protein data. This dual representation combines detailed local sequence information with global biochemical context that forms the core of the CVAE framework, enabling the generated proteins are both accurate in sequence and biologically plausible in their properties.

\subsection{BioAligner module}\label{subsec3}

To bridge the gap between textual annotations \( T \) and global physicochemical features \( \bar{f} \), we designed the BioAligner module, which aligns them in a shared latent space through contrastive learning. By mapping text annotations to biologically relevant physicochemical features, BioAligner ensures that semantic guidance from text can effectively influence protein generation. This alignment is crucial for tasks where text-based descriptions are needed to guide the design or generation of biologically active molecules, facilitating a robust and interpretable mapping between language and biological functionality.

In BioAligner, textual annotations are first encoded using a pre-trained model,  Sentence Transformer \cite{sentenceTransformer}, to generate embeddings \( t \). Simultaneously, global physicochemical features \( \bar{f} \), derived from protein sequences or domain-specific attributes (e.g., hydrophobicity and polarity), are processed through a transformer-based sequence encoder. These embeddings are then projected into a shared latent space where alignment occurs.

The alignment process is achieved by minimizing a contrastive loss function \(\L_{\text{contrastive}}\). Specifically, the process aims to minimize the distance between the embeddings of matching pairs \( (t, \bar{f}) \) while maximizing the distance between mismatched pairs \cite{contrastiveloss}.  The contrastive loss is defined in Equation\eqref{6}.

\begin{equation}
    L_{\text{contrastive}} = - \log \left( \frac{\exp(\text{sim}(t, \bar{f})/\tau)}{\sum_{i=1}^{N} \exp(\text{sim}(t, \bar{f}_i)/\tau)} \right) \label{6}
\end{equation}

 \( \text{sim}(t, \bar{f}) \) represents the cosine similarity between the embeddings, \( N \) is the batch size (or the number of negative samples), and \( \tau \) is a temperature parameter that controls the sharpness of the similarity distribution. This formulation encourages the embeddings of positive pairs to be pulled closer together in the latent space, while negative pairs are pushed apart. Additionally, the use of hard negatives, which are non-matching pairs that exhibit high cosine similarity, further improves alignment by focusing the model on challenging examples.

While batch sampling,  each batch contains both positive and negative pairs. Positive pairs consist of aligned \( T \) and \( \bar{f} \), while negative pairs are generated by mismatched text annotations with unrelated physicochemical features. Both types of embedding are optimized to improve  alignment in the shared space, and dynamic temperature adjustment helps fine-tune the convergence and flexibility of the model.

The BioAligner module processes text-guided physicochemical features \( \bar{f}_t \), which serve as conditioning inputs for subsequent diffusion models. These aligned features ensure that the semantic meaning of text annotations is faithfully translated into biologically relevant properties. This alignment enables BioAligner to effectively support applications where textual descriptions guide the design of biologically active molecules, improving the overall accuracy and interpretability of generated sequences.

\begin{figure}[h]
\centering
\includegraphics[width=3.5in]{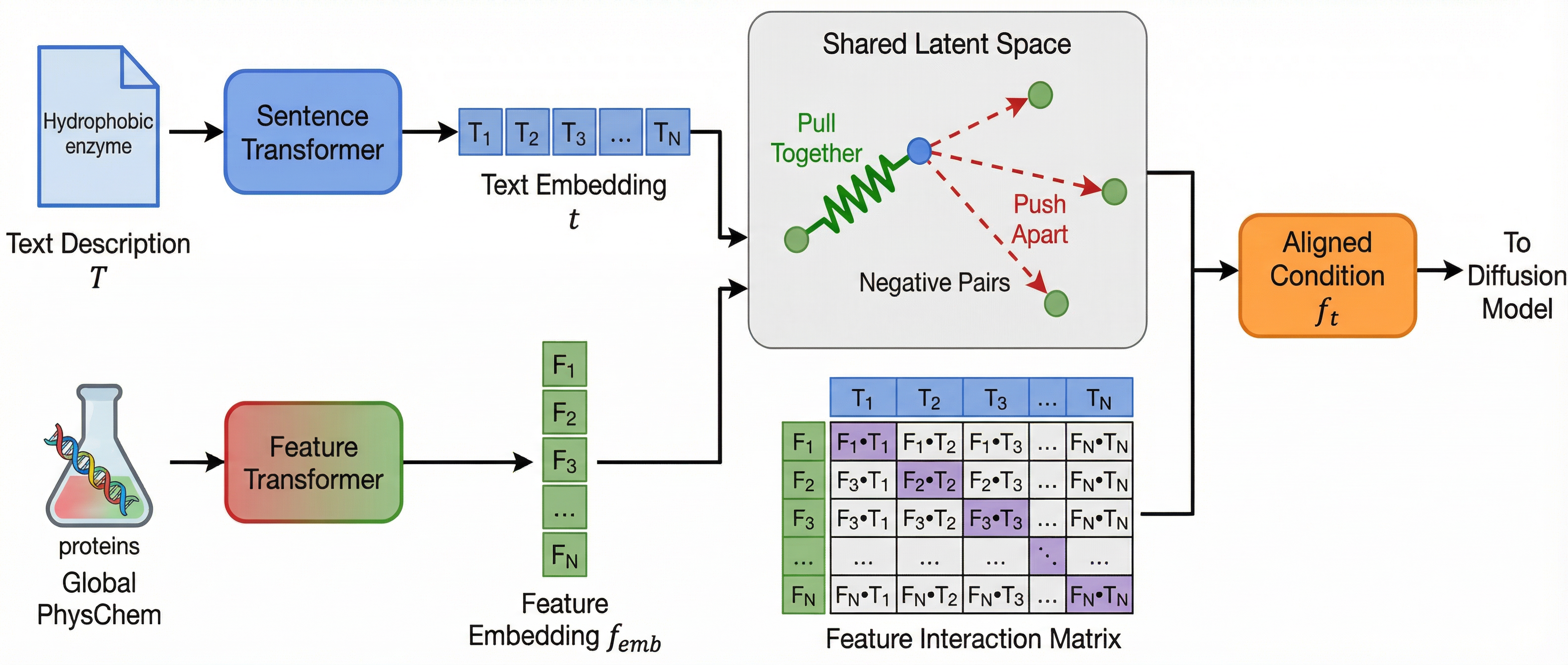}
\caption{BioAligner architecture}\label{bioaligner}
\end{figure}

In  alignment, as presented in \textbf{Figure} \ref{bioaligner}, BioAligner generates a matrix that represents the interactions between text and physicochemical features. This matrix is used to influence protein sequence generation, ensuring that the generated proteins are not only syntactically accurate, but also biologically relevant to the given textual descriptions.

\subsection{Text-property conditioned diffusion}\label{subsec4}
The conditional diffusion part generates protein sequences by iteratively denoising the latent state while incorporating guidance from physicochemical properties and text annotations. Text features, including biological and functional descriptions, are aligned with physicochemical properties through the BioAligner module. This alignment ensures that the generated sequences are biologically meaningful and adhere to the specified design requirements, maintaining global physicochemical coherence.

We adopt a diffusion probabilistic model to generate protein sequences conditioned on text-guided features \( \bar{f}_t \). The diffusion process adds noise to latent representations over \( T \) time steps progressively, transforming the original data into Gaussian noise \cite{ldm}. This process is modeled as Equation\eqref{7}.
\begin{equation}
    q(z_t | z_{t-1}) = \mathcal{N}(z_t; \sqrt{\alpha_t} z_{t-1}, (1 - \alpha_t) I)\label{7}
\end{equation}

 \( \alpha_t \) defines the noise schedule. The reverse process then denoises \( z_t \) to recover the original data while incorporating the conditioning features \( \bar{f}_t \). The denoising network \( f_{\text{denoise}} \) estimates \( z_{t-1} \) from \( z_t \) as Equation\eqref{8} shows.
\begin{equation}
p_{\theta}(z_{t-1} | z_t, \bar{f}_t) = \mathcal{N}(z_{t-1}; \mu_{\theta}(z_t, \bar{f}_t, t), \sigma_{\theta}^2 I) \label{8}
\end{equation}

\( \mu_{\theta} \) and \( \sigma_{\theta} \) are the predicted mean and variance. The conditioning features \( \bar{f}_t \) are integrated into the denoising network at each time step, which steers the generation process towards sequences with the desired semantic and biological properties.

\begin{figure}[h]
\centering
\includegraphics[width=3.5in]{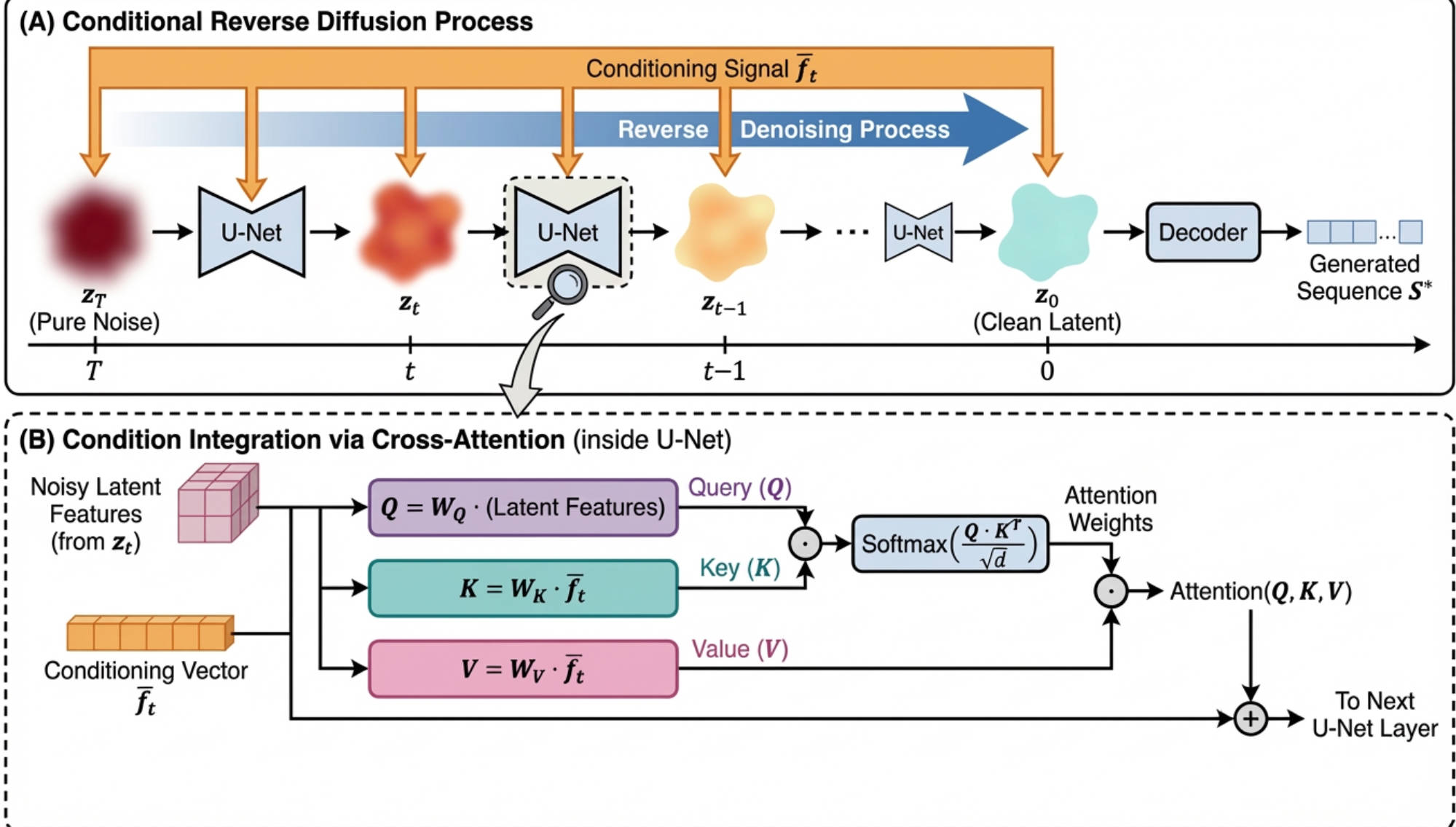}
\caption{Conditional diffusion architecture}\label{diff}
\end{figure}

To enable the effective integration of $\bar{f}_t$, text-guided features are extracted via the pretrained \textbf{BioAligner}, which maps semantic descriptions and physicochemical properties into a shared embedding space. These aligned features, $\bar{f}_t$, are subsequently injected into the latent diffusion process through a cross-attention mechanism within the denoising U-Net architecture. Specifically, the latent protein representations serve as the Queries ($Q$), while $\bar{f}_t$ provides the corresponding Keys ($K$) and Values ($V$). This configuration allows the model to dynamically assign attention weights to specific biological traits at each denoising step, thereby precisely steering the generation of protein sequences to align with the specified textual conditions and functional requirements, as illustrated in Figure 5.

The objective function minimizes the reconstruction error \(L_\text{diffussion}\) in the latent space as Equation\eqref{9}.

\begin{equation}
    L_{\text{diffusion}} = \mathbb{E}_{(t, z_t, \bar{f}_t)} \left[ \| z_t - \hat{z}_t \|^2 \right] \label{9}
\end{equation}

 \( \hat{z}_t \) is the denoised latent predicted by \( f_{\text{denoise}} \). The loss encourages the model to predict noise accurately while leveraging \( \bar{f}_t \) to condition the sequence generation process. The tight coupling of text conditions and the diffusion framework ensures that the output is biologically meaningful and aligns with the input text annotations semantically.

\section{Experiments}

To evaluate the functional relevance of protein sequences generated by CMADiff, an analysis of key physicochemical properties—hydrophobicity, isoelectric point (pI), net charge, and cysteine pair patterns—was conducted. These properties are essential for understanding protein folding, stability, and interactions.

\subsection{Data process}
To include both the protein sequence and its textual functional description, we chose the Swiss-Prot dataset for training. The Swiss-Prot dataset \cite{swissprot} is the expertly curated component of UniProt KB (produced by the UniProt consortium). The dataset contains hundreds of thousands of protein descriptions that have accurate annotation, including function, subcellular location, domain structure, post-translational modifications and functionally characterized variants.  These manual annotations mainly come from the research outputs of the literature and have been tested by the E-value. They are quite suitable for learning bioinformation-text-protein sequence representation. Thus, we built ProtSemantic dataset, a paired dataset of protein sequences and protein semantic context information, resorting to the Swiss-Prot dataset for high-quality protein annotations and constructing the property description of each protein. To construct the ProtSemantic dataset, we extracted and paired protein sequences with their corresponding semantic context information, using the following fields:

1. OC (organism classification): Describes taxonomic lineage (e.g. Eukaryota, Viridiplantae).

2. CC (comments and notes): Includes function, subcellular localization, and domain structure.

3. KW (keywords): Lists high-level terms related to protein properties, such as chloroplast, RNA-binding and ribosomal protein.

Each protein is represented by a structured description, which combines above fields to form high-quality semantic annotations. The pairing of sequences and annotations are jointed for training CMADiff, which generates  proteins from their functional contexts.

The physicochemical features dataset comprises  physical and chemical properties of 20 amino acids, including steric parameter (stc), helix probability (alpha), sheet probability (beta), hydrophobicity (H\_1), hydrophilicity (H\_2), polarity (P\_1), polarizability (P\_2), isoelectric pH (P\_i), side chain net charge number (NCN), solvent accessible surface area (SASA), accessibility (A1), antigenicity (A2), turn propensity (T), exposedness (E), and flexibility (F). The physicochemical properties integrated into CMADiff are selected based on their critical roles in protein folding and function. These features, as reported in  \textbf{Table F.1} of Appendix, capture a broad range of critical information that influences protein structure, stability, folding, and interactions.

\subsection{Evaluation metrics}
\label{sec_4_1}

We adopted the metrics shown in the \textbf{Table} \ref{tab:metrics} to evaluate the model. Relevant details are available in the Appendix.

\begin{table*}[htbp]
\centering
\footnotesize 
\caption{Summary of evaluation metrics}
\label{tab:metrics}
\begin{tabular}{|l|l|}
\hline
\textbf{Metric}                    & \textbf{Description}                                                    \\ \hline
\textbf{Physicochemical Property Analysis}    & Hydrophobicity, Electrostatic Properties, Disulfide Bond Patterns                                          \\ \hline
\textbf{Structural plausibility}    & pLDDT, TM-Score, RMSD, Fident                                          \\ \hline
\textbf{Functional relevance}       & Physicochemical Property Alignment, Textual Semantic Fidelity          \\ \hline
\textbf{Diversity and novelty}      & Sequence Shannon entropy, Novelty Ratio                          \\ \hline
\textbf{Computational efficiency}   & Generation Time per Sequence                                           \\ \hline
\end{tabular}

\end{table*}

\subsection{Training details }
The model architecture consists of CVAE and a DDPM model for sequence generation. The CVAE has a latent dimension of 512, a hidden dimension of 256, and a KL divergence weight of 0.5. The sequence and text encoders both use an embedding dimension of 384, with the text encoder based on the pretrained all-MiniLM-L6-v2 model. The DDPM model, built on a 1D U-Net architecture, is used for sequence generation through a denoising process. Training is conducted using the Adam optimizer with a learning rate of \( 1 \times 10^{-4} \), and the learning rate is adjusted using a StepLR scheduler with a decay factor of 0.9 every 30 epochs. The batch size is set to 64, with the CVAE trained for 500 epochs and the DDPM model trained for 100 epochs. 

\textbf{Hardware:} Experiments were conducted on an NVIDIA A100(80G) GPU cluster using PyTorch Lightning for training efficiency.

\textbf{Baselines:} 
We evaluate our model against several competitive baseline approaches, listed below: The Conditional Variational Autoencoder (CVAE), a classical probabilistic model, as a baseline for protein generation comparisons. The Left-to-Right Autoregressive (LRAR) and Convolutional Autoencoding Representations of Proteins (CARP) models \cite{CARP}, where both utilize dilated convolutional neural networks, trained on the UniRef50 dataset. The ProtGPT2 \cite{protgpt2}, an autoregressive protein language model based on GPT2 \cite{gpt2}, which has been pre-trained on the UniRef50 data as well. The EvoDiff, which leverages evolutionary-scale data within a diffusion model framework, focusing on sequence-first design. The TaxDiff, which integrates biological taxonomy information with the generative capabilities of diffusion models to generate structurally stable proteins within the sequence space. 
\subsection{Quantitative results}

\subsubsection{Physicochemical Property} 
To  investigate the functional relevance of the generated protein sequences, we analyzed key physicochemical properties, including hydrophobicity, isoelectric point (pI), net charge, and cysteine pair patterns(disulfide bond). These properties are critical for understanding protein folding, stability, and interactions. The analysis compared proteins generated by CMADiff with natural proteins and those generated by baseline models (TaxDiff and EvoDiff).

\begin{figure}[h]
\centering
\includegraphics[width=3.5in]{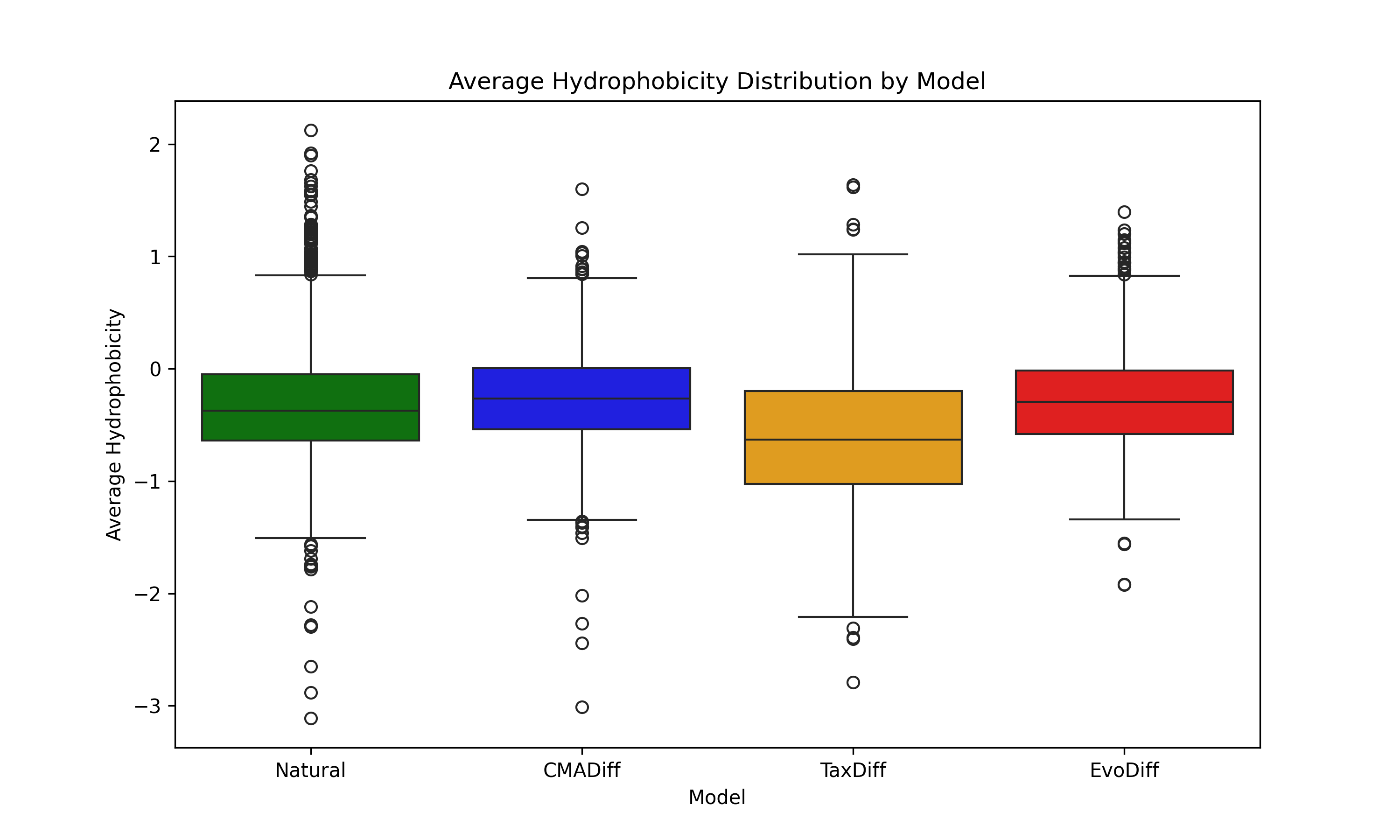}
\caption{Average Hydrophobicity Distribution by Model}\label{avg_h}
\end{figure}

 CMADiff excels in generating sequences with hydrophobicity values that closely match natural proteins, with mean and median values of -0.28 and -0.27, as illustrated  in \textbf{Figure} \ref{avg_h}. This indicates a similar distribution of hydrophobic and hydrophilic residues to natural proteins. In contrast, TaxDiff produces sequences with significantly lower hydrophobicity (mean = -0.63, median = -0.63), suggesting a tendency to generate more hydrophilic sequences, which may affect structural stability. EvoDiff's hydrophobicity values are closer to CMADiff (mean = -0.27, median = -0.30) but show less consistency with a higher standard deviation (0.46 vs. 0.44).

The relationship between the isoelectric point (pI) and net charge at pH 7.4  in \textbf{Figure} \ref{pi}  and  pI distribution in appendix highlights how well each model captures the charge properties of proteins. Regarding charge properties, CMADiff closely mirrors the natural distribution of pI and net charge, indicating its ability to generate proteins with realistic charge characteristics. Natural proteins show a clear correlation between pI and net charge, with proteins above a pI of 7.4 being positively charged and those below being negatively charged. CMADiff's generated proteins exhibit this same correlation, while TaxDiff shows a narrower range of pI and net charge values with a less defined correlation. EvoDiff displays a broader range of net charge values but with less consistency in the correlation to pI.

In terms of cysteine pair intervals as shown in \textbf{Figure} \ref{cys}, which are crucial for disulfide bond formation and protein structure stabilization, CMADiff and natural proteins exhibit a broad and similar distribution. This indicates the ability to form diverse disulfide bond patterns. TaxDiff shows a significantly narrower distribution, suggesting limited variability in disulfide bond formation, which may restrict structural diversity. EvoDiff also exhibits a narrower distribution compared to CMADiff, though slightly broader than TaxDiff, indicating some limitations in capturing the full range of disulfide bond patterns.

Overall, CMADiff demonstrates a strong capacity to generate proteins with functional and structural characteristics closely resembling natural proteins. This success is largely attributed to its Conditional Variational Autoencoder (CVAE) module, which effectively integrates local and global physicochemical properties. By combining these properties, the model captures both detailed sequence information and overall biochemical context, ensuring structurally plausible and functionally relevant proteins.
In contrast, TaxDiff and EvoDiff show notable limitations. TaxDiff, relying primarily on taxonomic information, lacks detailed physicochemical insights, resulting in a narrower range of pI, net charge, and limited disulfide bond variability. EvoDiff, while generating sequences with a broader net charge range, exhibits inconsistency in pI-charge correlation and less balanced hydrophobicity distribution, indicating incomplete capture of physicochemical relationships.
CMADiff's superior performance highlights the importance of integrating both local and global physicochemical properties, enabling precise and controlled protein design that closely mimics natural proteins.

\begin{figure}[h]
\centering
\includegraphics[width=3.5in]{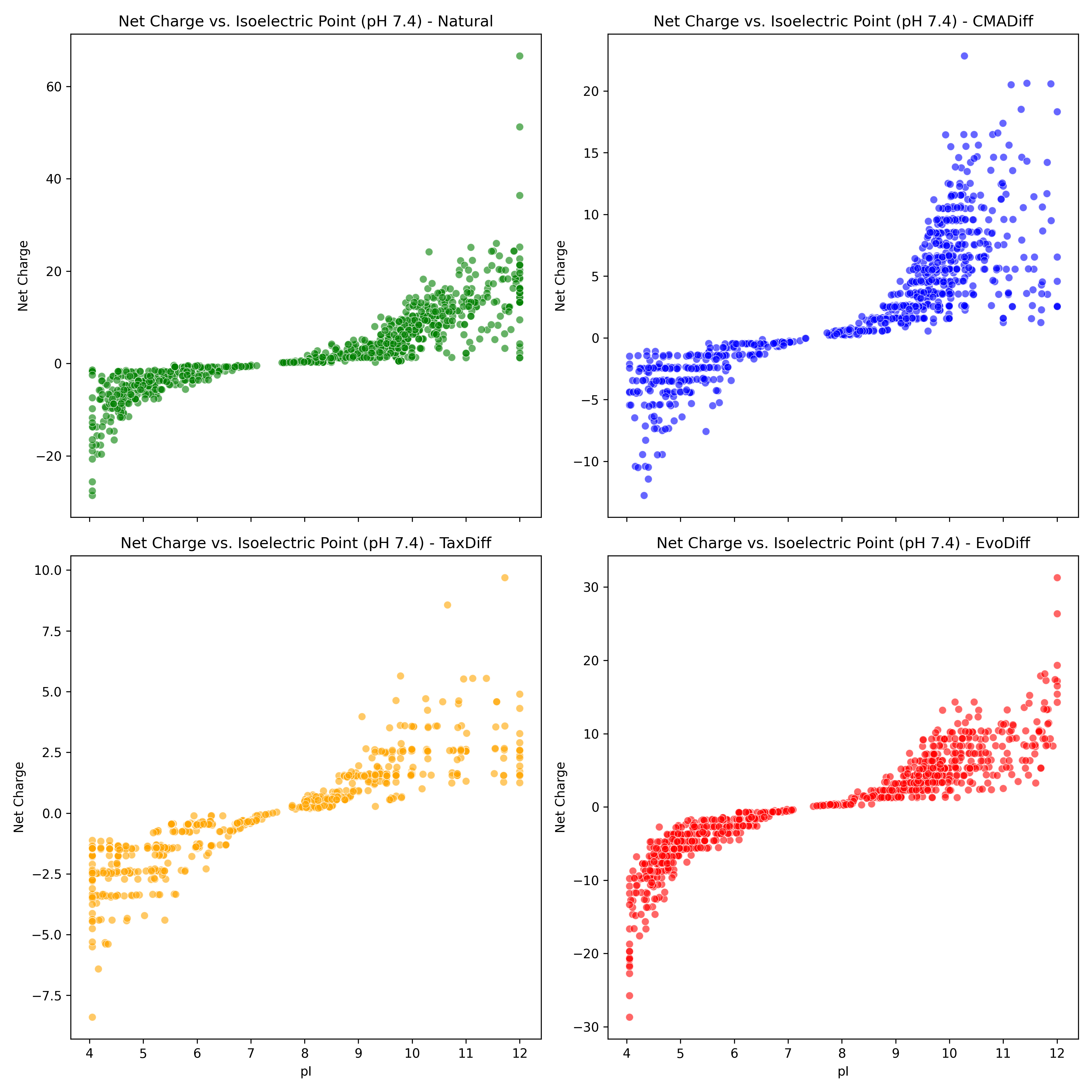}
\caption{Net Charge vs. Isoelectric Point (pH 7.4)}\label{pi}
\end{figure}

\begin{figure}[h]
\centering
\includegraphics[width=3.5in]{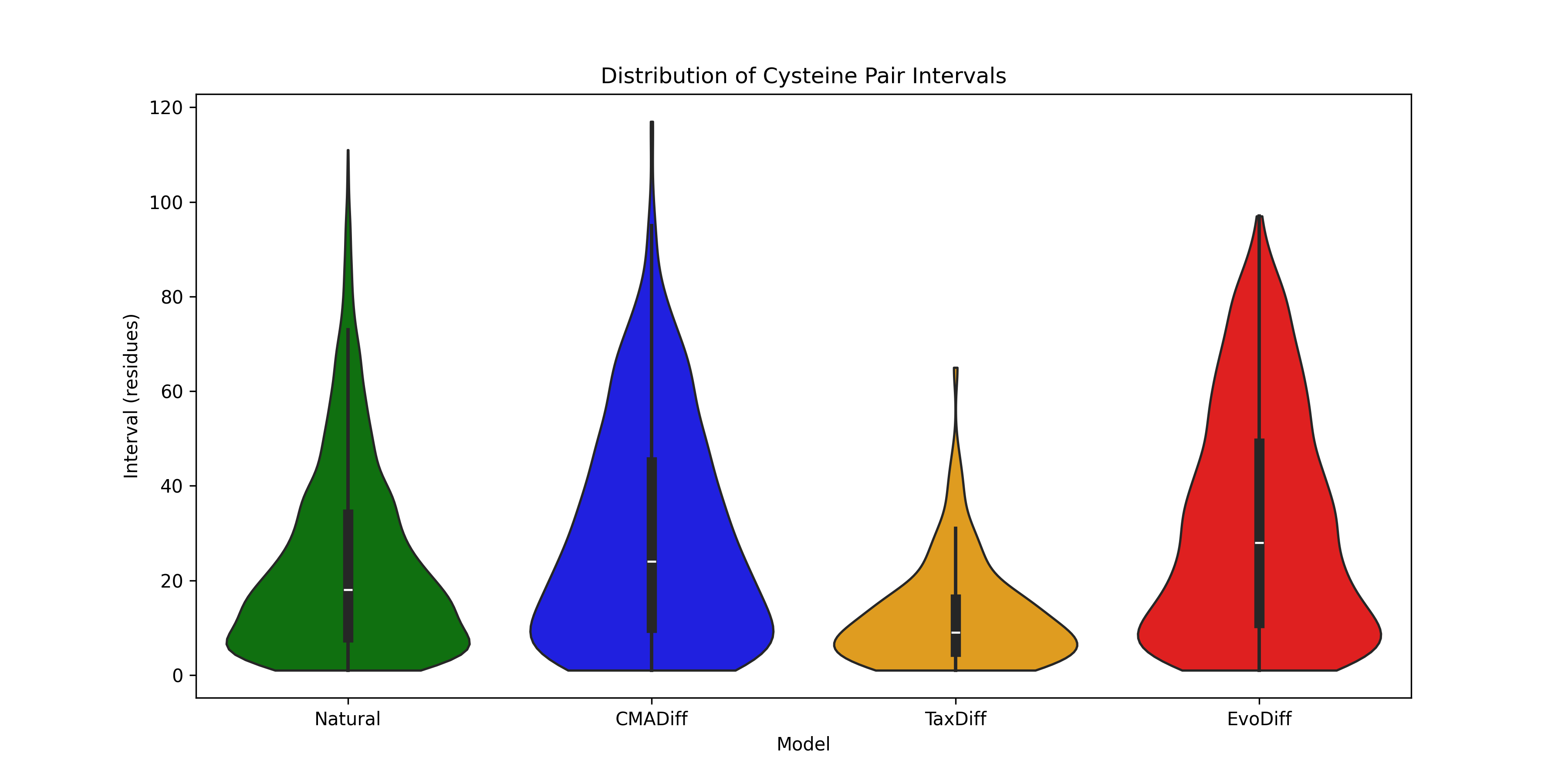}
\caption{Distribution of Cysteine Pair Intervals(disulfide bond)}\label{cys}
\end{figure}

\subsubsection{Structural plausibility} 

From \textbf{Table \ref{table:1}}, it is evident that CMADiff outperforms all baseline models in terms of pLDDT, TM-score, RMSD, and Fident. When using raw features, CMADiff achieves superior structural quality, exhibiting enhanced stability and topological similarity compared to the other models.The superior performance of CMADiff in structural plausibility can be ascribed to the integration of protein sequences and physicochemical properties through the CVAE module, which forms a robust latent space capturing biological traits and ensuring the generation of structurally plausible sequences. Additionally, the conditional diffusion process, guided by BioAligner, refines the noisy latent representations to generate structurally accurate protein sequences by integrating physicochemical properties and text annotations, ensuring that the generated sequences meet the desired structural criteria.

\begin{table*}[htb]
\centering
\footnotesize
\setlength{\tabcolsep}{0.2mm}  
{
\caption{\textbf{Controllable generation on Alphafold/Swiss-Prot and PDB datasets}}
\label{table:1}
\vskip 0.1in
{
\begin{tabular}{llc|ccc|ccc}
\toprule
\multirow{2}{*}{\textbf{Method}} & \multirow{2}{*}{\textbf{pLDDT}} & \multirow{2}{*}{\textbf{Time(mins)}} & \multicolumn{3}{c|}{\textbf{Swiss-Prot Dataset}} & \multicolumn{3}{c}{\textbf{AFDB Dataset}} \\
\cmidrule(rl){4-6}\cmidrule(rl){7-9}
 &  &  & \textbf{TM-score} & \textbf{RMSD} & \textbf{Fident} & \textbf{TM-score} & \textbf{RMSD} & \textbf{Fident} \\
 &  &  & \textbf{(\%)}$\uparrow$ & \textbf{(\(\AA\))}$\downarrow$ & \textbf{(\%)}$\uparrow$ & \textbf{(\%)}$\uparrow$ & \textbf{(\(\AA\))}$\downarrow$ & \textbf{(\%)}$\uparrow$ \\
\midrule
natural & 79.45 ± 13.48 & / & 59.03 & 0.45 & 23.77 & 75.26 & 0.32 & 37.67 \\
CARP & 45.94 ± 12.78 & 95.21 & 39.84 & 12.65 & 11.93 & 34.52 & 11.82 & 11.78 \\
LRAR & 46.32 ± 13.15 & 80.15 & 38.16 & 12.57 & 15.89 & 29.18 & 13.45 & 16.35 \\
ProtGPT2 & 51.78 ± 15.69 & 37.58 & 40.05 & 12.77 & 11.59 & 31.11 & 12.85 & 13.58 \\
Evodiff & 51.57 ± 11.94 & 35.13 & 48.13 & 11.26 & 14.77 & 48.70 & 12.14 & 15.09 \\
Taxdiff & 68.66 ± 9.81 & 25.14 & 49.77 & 4.84 & 17.78 & 51.07 & 4.89 & 15.30 \\
\textbf{CMADiff} & \multirow{2}{*}{\textbf{70.57 ± 14.13}} & \multirow{2}{*}{\textbf{13.59}}  & \multirow{2}{*} {\textbf{52.04}} & \multirow{2}{*}  {\textbf{4.28}} & \multirow{2}{*}  {\textbf{17.63}} & \multirow{2}{*} {\textbf{53.62}} & \multirow{2}{*}  {\textbf{4.85}} & \multirow{2}{*}  {\textbf{17.88}} \\
\textbf{(raw\_feature)} & & & & & & & & \\
\textbf{CMADiff} & \multirow{2}{*}{\textbf{55.14 ± 14.79}} & \multirow{2}{*}{\textbf{13.35}}& \multirow{2}{*} { \textbf{47.38}} & \multirow{2}{*}  {\textbf{7.26}} & \multirow{2}{*}  {\textbf{14.75}} & \multirow{2}{*}  {\textbf{48.57}} & \multirow{2}{*}  {\textbf{8.15}} & \multirow{2}{*}  {\textbf{15.16}} \\
\textbf{(random\_feature)} & & & & & & & & \\
\bottomrule
\end{tabular}
}
}
\vskip -0.1in
\begin{flushleft}
\small \textbf{Note:} Metrics are calculated with 1000 samples generated from each model, with lengths following a random distribution between 10 and 128. The sampling time was recorded on a single GPU for 1000 samples.
\end{flushleft}
\end{table*}

\subsubsection{Functional relevance}

\begin{figure}[h]
\centering
\includegraphics[width=3.5in]{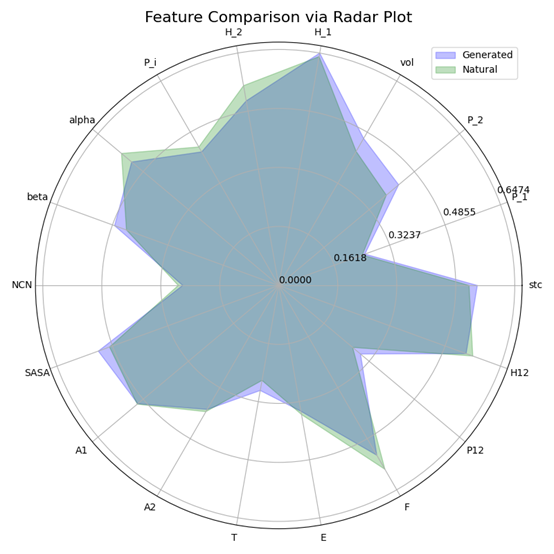}
\caption{The Feature Distribution Comparision between Generated and Natural Sequences}\label{feature}
\end{figure}

From \textbf{Figure \ref{feature}}, it can be observed that the distributions of generated and natural sequences are similar in shape, location, and width across most features, indicating that the model effectively simulates the distribution of natural sequences in these physicochemical properties. The closeness of their medians further emphasizes the functional relevance of the model. This also verifies the effectiveness of the CVAE encoder in mapping data to the latent space, as the training result shows in appendix. In \textbf{Figure \ref{matrix}}, high similarity along the diagonal further supports the successful alignment between matching text and sequences. Lower similarity in off-diagonal areas indicates the  ability to distinguish non-matching pairs, demonstrating the strong text-sequence alignment capabilities of BioAligner. Additionally, a high Textual Semantic Fidelity scores is an indicator of strong alignment between input descriptions and generated sequences. With CMADiff achieving a semantic similarity score of 0.94, this result emphasizes the effectiveness of the BioAligner module in maintaining semantic relevance.

\begin{figure}[h]
\centering
\includegraphics[width=2.5in]{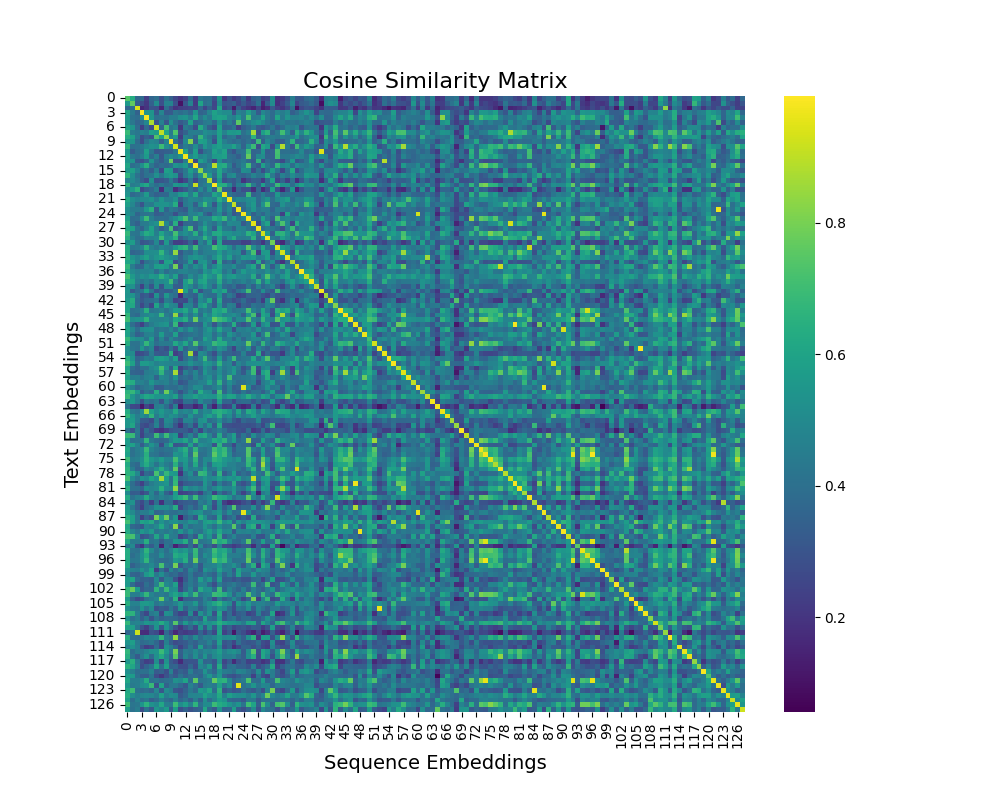}
\caption{Similarity matrix of physicochemical and text description vectors in BioAligner}\label{matrix}
\end{figure}

\subsubsection{Computational efficiency analysis}

CMADiff took an average of 13.59 minutes for generating 1,000 sequences, which is significantly faster than other models showcasing its computational advantages. The computational efficiency of CMADiff can be assigned to its efficient model architecture, which enables rapid processing of protein sequences and physicochemical features.

\subsubsection{Diversity and novelty} 

The Sequence Shannon entropy reflects the variability in amino acid composition across all generated sequences, with higher values indicating broader exploration of the sequence space. The shanno entropy comparation with natural sequence and random sequence is shown in appendix. This suggests that CMADiff is particularly effective in exploring diverse sequence spaces, which is critical for  protein design.

To assess novelty, protein sequences with an average pLDDT over 70 were evaluated against the PDB database using stringent criteria (E-value \textgreater 1e-5, Identity \textless 20\%). CMADiff achieved a Novelty Ratio of 0.78, outperforms TaxDiff  and EvoDiff. This demonstrates its strong capability to generate novel protein sequences with minimal overlap with existing databases, enabling the discovery of unique protein structures and functions.

The elevated diversity and originality of CMADiff can be ascribed to the incorporation of both local and global physicochemical characteristics. This integration enables the model to traverse a vast array of sequence spaces and yield a plethora of protein sequences. Moreover, the conditional diffusion process, directed by text annotations and physicochemical properties, guarantees that the generated sequences are not only diverse but also functionally pertinent and structurally plausible.

\subsection{Ablation study}

In ablation study, we evaluated the contribution of each component in the CMADiff model. First, removing the conditional diffusion process led to a significant drop in pLDDT score, showing its critical role in improving protein structure quality. Another trial is to remove the BioAligner module, where the pLDDT score declined, despite other metrics improved moderately, demonstrating its importance in aligning sequences with physicochemical properties and textual descriptions. The full CMADiff model, especially when using raw features, outperformed all others, showing that raw features retain essential biological information which enhances structural and topological similarity. These findings exhibit a great potential of our model for protein engineering and drug discovery, as shown in \textbf{Table \ref{table:2}}.

\begin{table*}[htb]
\centering
\footnotesize
\setlength{\tabcolsep}{0.3mm}  
{
\caption{Ablation results of different parts on Alphafold/Swiss-Prot and PDB datasets. }
\label{table:2}
\vskip 0.1in
{
\begin{tabular}{cl|ccc|ccc}
\toprule
 \multirow{2}{*}{\textbf{Part}} 
 & \multirow{2}{*}{\textbf{pLDDT$\uparrow$}}  
 & \multicolumn{3}{c|}{\textbf{AFDB Dataset}} & \multicolumn{3}{c}{\textbf{PDB Dataset}} \\
 \cmidrule(rl){3-5}\cmidrule(rl){6-8}
  &  &  \textbf{TM-score} & \textbf{RMSD} & \textbf{Fident} & \textbf{TM-score} & \textbf{RMSD} &\textbf{ Fident} \\
  &  & \textbf{(\%)}$\uparrow$ & \textbf{(\(\AA\))}$\downarrow$ & \textbf{(\%)}$\uparrow$ & \textbf{(\%)}$\uparrow$ & \textbf{(\(\AA\))}$\downarrow$ & \textbf{(\%)}$\uparrow$ \\
\midrule
only CVAE  & 37.69 ± 9.47 & 40.05 & 13.25 & 11.54 & 36.37 & 12.36 & 11.25  \\
without BioAlinger  & 49.07± 15.21 & 42.37 & 10.54 & 12.75 & 43.57 & 10.46 & 12.14  \\
\textbf{Whole(raw\_feature)} & \textbf{70.57 ± 14.13} & \textbf{52.04} & \textbf{4.28} & \textbf{17.63} & \textbf{53.62} & \textbf{4.85} & \textbf{17.88} \\
\textbf{Whole(random\_feature)} & \textbf{55.14 ± 14.79}  & \textbf{47.38} & \textbf{7.26} & \textbf{14.75} & \textbf{48.57} & \textbf{8.15} & \textbf{15.16} \\
\bottomrule
\end{tabular}
}
\vskip -0.1in
\begin{flushleft}
\small \textbf{Note:} Metrics are calculated with 1000 samples generated from each model.
\end{flushleft}
}
\end{table*}

\section{Use cases}
In order to  show the effect of our model more intuitively, we conducted 
 qualitative analysis using AlphaFold 3 \cite{AlphaFold3}. Below are two examples.

\subsection{Case study 1}
\textbf{The original physical and chemical properties of nature}
\\ \textbf{Input}:\\
\\
\textbf{Species}: Eukaryota;  Metazoa;  Spiralia;  Lophotrochozoa;  Mollusca;  Gastropoda; Caenogastropoda;  Neogastropoda;  Conoidea;  Conidae;  Conus;  Pionoconus.\\
\\
\textbf{Keyword}:Cleavage on pair of basic residues;  Disulfide bond;  Hydroxylation; Ion channel impairing toxin;  Knottin;  Neurotoxin;  Presynaptic neurotoxin; Secreted;  Signal;  Toxin;  Voltage-gated sodium channel impairing toxin.
\\
\\
\textbf{Function}: Delta-conotoxins bind to site 6 of voltage-gated sodiumchannels (Nav) and inhibit the inactivation process. SUBCELLULAR LOCATION: Secreted.  TISSUE SPECIFICITY: Expressed by the venom duct. DOMAIN: The presence of a disulfide through disulfide knot'structurally defines this protein as a knottin. DOMAIN: The cysteine framework is VI/VII (C-C-CC-C-C).  SIMILARITY: Belongs to the conotoxin O1 superfamily.\\

\textbf{Output}:MKLTCVVIVAVLFLTAWTFVMADDSRYGLKDLFPKARHEMKNPES\\KLNKRDECFSPGTFCGIKPGLCCSAWCYSFFCLTLTF\\

To evaluate the proposed framework, we generated protein sequences based on specified textual descriptions and physicochemical properties. The AlphaFold 3 predictions of the 3D structures in \textbf{Figure \ref{AlphaFold}}(a) showed consistently high pLDDT scores, indicating reliable structural confidence. 
In addition, Foldseek\cite{foldseek} analysis in \textbf{Figure \ref{AlphaFold}}(b) demonstrated high TM-scores, low RMSD values, and species consistency with natural homologs from databases like Swiss-Prot. These results confirmed that CMADiff can effectively design structurally reliable and biologically relevant proteins, while aligning with input constraints and species-specific requirements.\subsection{Case study 2}
\textbf{Stochastic new physicochemical properties}
\\ \textbf{Input}: Random physicochemical properties
\\ \textbf{Output}:KGWNLRKKARENTIQFINFWDCVREYTERKHNE

\begin{figure}[H]
\centering
\includegraphics[width=3.5in]{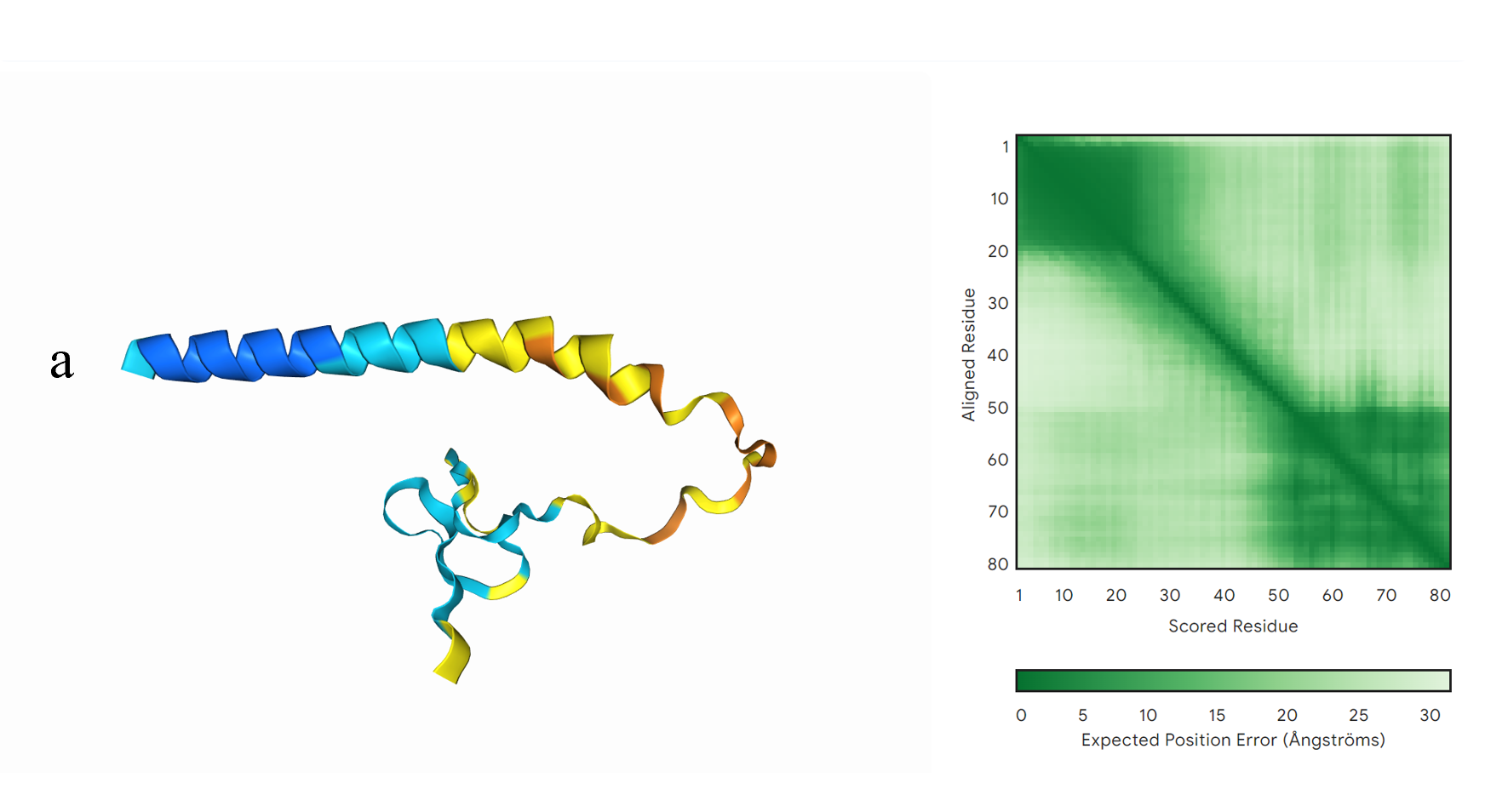}  
\includegraphics[width=3.5in]{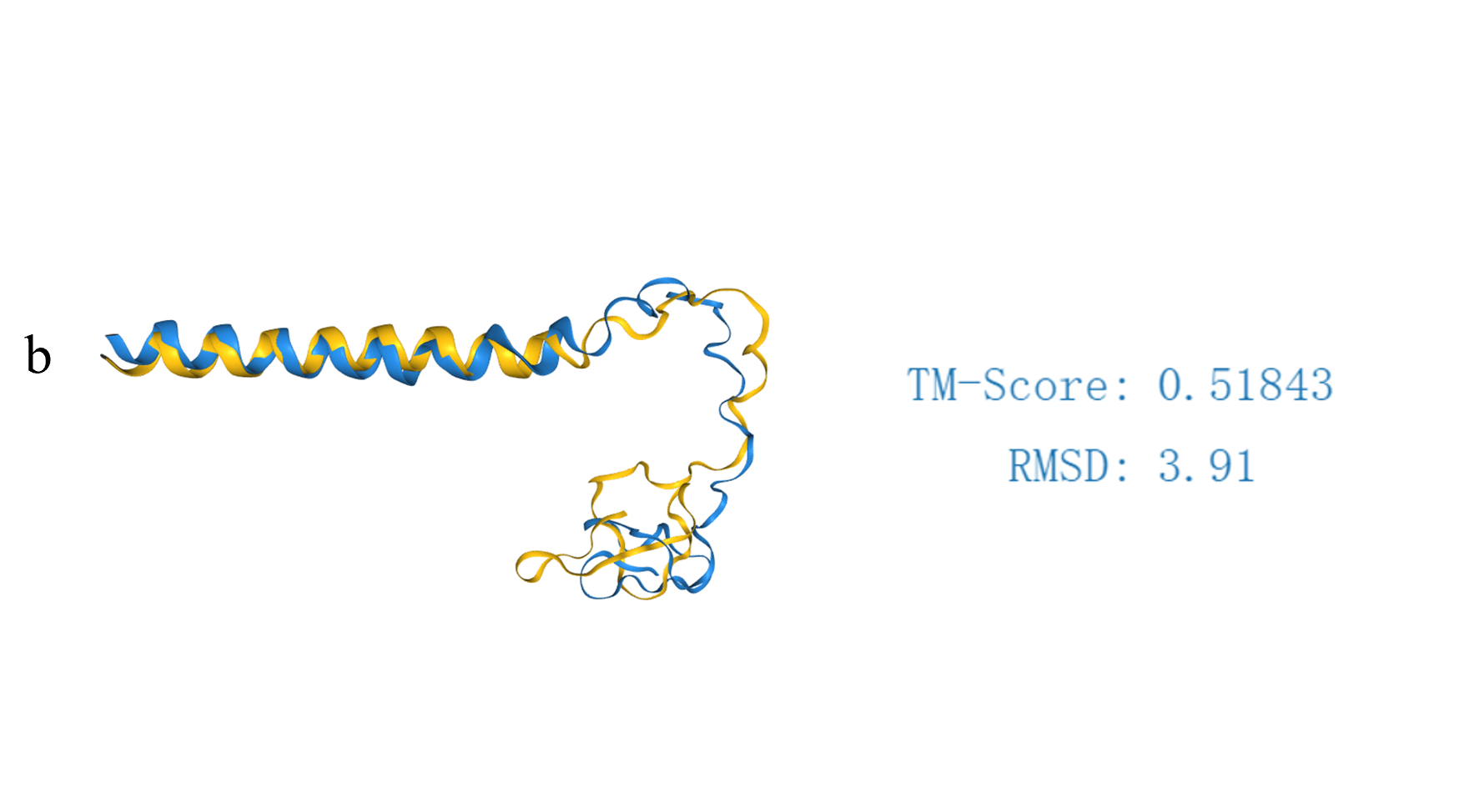}
\includegraphics[width=3.5in]{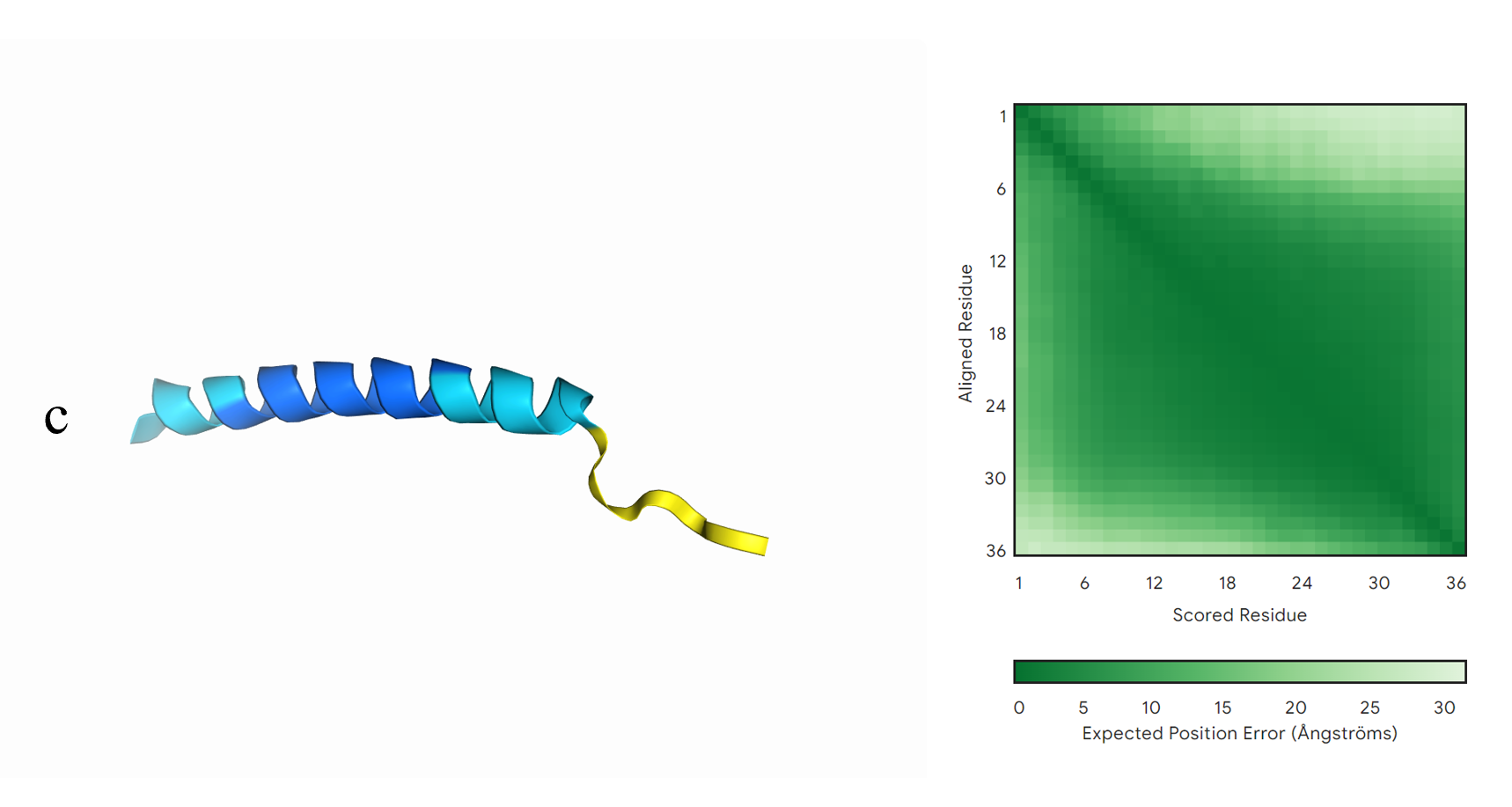}  
\caption{These three pictures are the results of using AlphaFold and FoldSeek to evaluate the case. \textbf{(a)} Using original text to describe the score of the generated protein in AlphaFold 3 in Case 1. \textbf{(b)} The sequence generated in Case 1 was aligned with structurally similar native proteins in Foldseek. \textbf{(c)} The proteins generated in Case 2, which exhibit random physical and chemical properties, demonstrate high plDDT scores.}
\label{AlphaFold}
\end{figure}

We further explored the  capacity of the model by generating protein sequences under randomly assigned physicochemical property conditions. Among the generated sequences, some achieved exceptionally high pLDDT scores (90+) when evaluated using AlphaFold 3 as shown in \textbf{Figure \ref{AlphaFold}}(c) , indicating high structural confidence. Notably, a subset of these sequences could not be matched to any existing proteins in natural databases when analyzed with Foldseek, suggesting that the model successfully generated novel protein sequences not observed in nature. This emphasizes the framework's potential for de novo protein design, showcasing its ability to generate unique sequences with reliable structural characteristics.

\section{Conclusion}

In this study, we presented \textbf{CMADiff}, a novel cross-modal aligned diffusion framework that bridges the gap between protein physicochemical properties and natural language descriptions. By integrating a CVAE-based latent space with the BioAligner module, CMADiff enables intuitive and controllable protein sequence generation.

\subsection{Evaluation of the Work}
The primary strength of CMADiff is its ability to generate sequences that are both structurally plausible  and functionally consistent with textual prompts. Its high efficiency and novelty ratio  demonstrate its potential for \textit{de novo} design. However, a limitation remains in its dependence on the quality of existing database annotations. Currently, the model focuses on primary sequences, with less emphasis on complex protein-protein interactions and dynamic conformational changes.

\subsection{Impact and Benefits to the Field}
CMADiff provides a transformative tool for protein engineers and bioinformaticians by offering a "text-to-protein" interface that significantly reduces the cost of experimental screening. By aligning semantic intent with biological traits, our framework empowers researchers to explore uncharted sequence spaces with high precision. Furthermore, the open-sourcing of our model and dual-modality alignment methodology provides a reusable blueprint for developing more interpretable and domain-aware generative AI in the broader life sciences community.

\subsection{Future Work}
Future research will focus on expanding functional diversity by incorporating multi-omics data and larger unlabeled datasets, while simultaneously embedding 3D geometric constraints directly into the diffusion process to enhance spatial accuracy. We also aim to conduct wet-lab synthesis to verify the biological activity of generated sequences and develop an interactive system for iterative design refinement via natural language feedback.
 
 Ultimately, CMADiff establishes a robust foundation for integrating biological domain knowledge into generative AI, paving the way for a more precise era of biotechnology applications.

\section*{Acknowledgments}
This work was supported by  the National Key Research and Development Program of China (no. 2023YFD1700700), and the Basic Research Support Program for Excellent Young Teachers in Heilongjiang Provincial Undergraduate Universities (no. YQJH2023202).

\section*{Declarations}

The authors declare that they have no known competing financial interests.

\section*{Data and Software Availability}
The implementation of CMADiff with detailed instructions on their contents and usage is  available at~\url{https://github.com/HPC-NEAU/CMADiff} under the MIT License. In addition, the Swiss-Prot dataset used in this study is publicly available via the UniProt Knowledgebase (UniProtKB)~\url{https://www.uniprot.org/uniprotkb}. Furthermore, the ProtSemantic dataset built by us  are  available at~\url{https://huggingface.co/sanyier312/CMADiff/tree/main}.
Structural evaluations were performed using AlphaFold3  available at~\url{ https://github.com/google-deepmind/alphafold3} and Foldseek~\url{https://github.com/steineggerlab/foldseek}.

\nocite{*}
\bibliographystyle{IEEEtran}
\bibliography{refs}

\end{document}